# COST CA20120 INTERACT Framework of Artificial Intelligence Based Channel Modeling

Ruisi He[1], *Senior Member, IEEE,* Nicola D. Cicco[2], *Member, IEEE,* Bo Ai[1], *Fellow, IEEE,* Mi Yang[1], *Member, IEEE,* Yang Miao[3], *Member, IEEE,* and Mate Boban[4], *Member, IEEE*

[1]School of Electronics and Information Engineering, Beijing Jiaotong University, Beijing, China
[2]Department of Electronics, Information, and Bioengineering, Politecnico Di Milano, Italy
[3]Electrical Engineering, Mathematics and Computer Science, University of Twente, Netherlands
[4]Huawei Technologies Duesseldorf GmbH, Munich Research Center, Germany

***Abstract*—Accurate channel models are the prerequisite for communication-theoretic investigations as well as system design. Channel modeling generally relies on statistical and deterministic approaches. However, there are still significant limits for the traditional modeling methods in terms of accuracy, generalization ability, and computational complexity. The fundamental reason is that establishing a quantified and accurate mapping between physical environment and channel characteristics becomes increasing challenging for modern communication systems. Here, in the context of COST CA20120 Action, we evaluate and discuss the feasibility and implementation of using artificial intelligence (AI) for channel modeling, and explore where the future of this field lies. Firstly, we present a framework of AI-based channel modeling to characterize complex wireless channels. Then, we highlight in detail some major challenges and present the possible solutions: i) estimating the uncertainty of AI-based channel predictions, ii) integrating prior knowledge of propagation to improve generalization capabilities, and iii) interpretable AI for channel modeling. We present and discuss illustrative numerical results to showcase the capabilities of AI-based channel modeling.**

## I. INTRODUCTION

Wireless channel modeling represents a fundamental challenge in development of wireless communications, and is of paramount importance for the design and performance evaluation of next-generation wireless networks. The growing number and density of devices, coupled with high-mobility users and increased traffic demands from emerging use cases in 5G and 6G, pose significant challenges in obtaining timely and reliable channel information to ensure optimal communication performance [1] [2]. The expansion of mobile communication systems in space-time-frequency domains has led to increased complexity in radio propagation, and made accurate channel modeling more challenging. The current channel modeling methods exhibit inherent limitations, failing to satisfy the demands of next-generation communication systems for high-precision generalizable channel models. As the propagation process of radio waves grows increasingly complex, the traditional statistical and deterministic channel models face increased complexity, resulting in constraints on the adaptability to various scenarios and capacity for self-evolution. Consequently, there is a pressing need to explore novel modeling methods and technologies capable of effectively addressing the intricacy and uncertainty of wireless channels. The advancement of artificial intelligence (AI) technology represents a promising avenue for revolutionizing wireless channel modeling. The deployment of massive communication nodes generates a wealth of data related to their respective channels and overall communication activities. This data offers opportunities for learning underlying patterns, which can be harnessed for efficient channel modeling and prediction.

AI has proven to be a powerful tool for modeling complex non-linear functions. For channel modeling and prediction, the relation between the channel to be modeled/predicted and the known/collected information about the channels also exhibits such non-linearity. To that end, a lot of work leveraged AI models, specifically Deep Neural Networks (DNNs), to learn mappings between collected measurements and the wireless channel [3][4]. On the one hand, when dealing with intricate channel modeling problems where the understanding of physical propagation environment is limited, an AI-based propagation model functions essentially as a black box: it identifies the underlying patterns between input data or output labels and the corresponding input features without providing a human-understandable explanation. On the other hand, when there are some inferred physical or theoretical insights available, AI can be directed toward enhancing the accuracy of a baseline propagation model by introducing effective correction factors.

In terms of the channel modeling task at hand, recent research mainly explores the use of AI in the following three broad directions: i) classification; ii) clustering; and iii) regression. In the context of channel modeling, classification can refer to, for instance, scenario identification [5]. Channels can be classified into different categories based on the collected channel features (e.g., high/low speeds and Doppler) or based on geographical or context information (e.g., highway, tunnel, urban). Clustering refers to the identification and grouping of multipath contributions with similar characteristics [6]. Multipath clustering plays a crucial role in simplifying the complexity of channel modeling while capturing the fundamental aspects of propagation. Finally, regression can help in the direct prediction of the channel characteristics by establishing the cause-and-effect relationship between propagation characteristics (such as received signal strength, path loss, spread coefficients of channel, etc.) and relevant input features associated with the propagation process for a specific channel sample [7]. The scope of AI-based channel modeling research is quite broad and encompasses more topics than can be realistically addressed in one single paper. For this reason, the main focus of the rest of this paper is on efficient and accurate regression problems for channel modeling and prediction. In other words, we focus on the modeling and prediction of channel characteristics - either in its full form in terms of channel impulse response or in terms of channel parameters such as path loss and delay/angle spreads.

For the modeling and prediction of radio channel, it is challenging to realize the mapping of complex environments and channel characteristics through neural network. The current methods mainly tend to be data-driven and lack of in-depth investigation on the design and architecture of the model. As a result, the interpretability, expansibility and generalization of the existing methods are often limited. Therefore, it is necessary to further explore the optimizied model to improve performance in complex wireless environments. Some potential solutions include: uncertainty-aware AI modeling, physics-informed AI modeling, and explainable AI modeling. They provide effective support for the generalization ability, reliability and interpretation of the model. General AI models often provide predictions without quantifying uncertainty, which can affect credibility in complex environments. Some existing methods improve the model robustness in the cases of scarce data or high noise, but it still needs to provide more accurate confidence intervals for prediction. Traditional AI models face limitations in capturing physical characteristics of wireless propagation channels. Physics-informed AI modeling can address this issue by incorporating physical laws or prior knowledge into the model, improving the generalization capabilities while reducing the dependence on large amounts of training data. In addition, the "black-box" nature of deep learning models limits the applicability in scenarios that require transparency and interpretability. In this paper, the existing problems for AI-based channel modeling are explored, and some specific contributions of this paper can be summarized as follows: i) We propose a general framework for AI-based channel modeling and prediction; ii) We explore the soultions to improve performance of AI-based channel modeling in complex wireless environment, and focus on model-driven design and the use of channel prior knowledge; iii) We make tentative explorations of several key directions. A combined quantile regression with conformal prediction techniques for uncertainty modeling method is presented, a physics-enhanced model is designed, and a symbolic regression modeling method is proposed to provide a global explanation of the relationship between input and output.

This paper is an outcome of the work initially performed within European Commission-Funded COST CA20120 Action Intelligence-Enabling Radio Communications for Seamless Inclusive Interactions (COST INTERACT) [8]. The main goal of COST INTERACT is to go beyond the capabilities of current networks to make the radio network itself intelligent. Within COST INTERACT, the Working Group 1 (WG1) deals with theoretical and experimental understanding of radio channels and at deriving models for design, simulation, and planning of future wireless systems. Recently, WG1 has finalized its Whitepaper on channel modeling for future communications systems [9], where AI-based channel modeling was featured prominently. Stemming from that initial work, this paper delves deep in the topic by describing a framework and indicating the main directions for AI-based channel modeling.

## II. General Framework of AI-Based Channel Modeling

Complex and dynamic environments and radio propagation processes have made it increasingly challenging to mine and predict the underlying features of massive channel data. Developments of AI technologies have laid the foundation for establishing deep and complex correlations between environmental information and propagation mechanisms. AI can effectively use environmental information and historical data to learn and extract complex nonlinear features in channels, thereby enabling accurate prediction of channel characteristics. Here, we present the proposed general framework of AI-based channel modeling as in Fig. 1. The framework is composed of input, output, and AI model components. The input generally includes channel- and environment-related data. Channel-related data includes raw channel data and various channel features, whereas environment-related data includes spatial data (e.g., satellite images), quantitative data (e.g., building density), and temporal data (e.g., hydrological data). The output of AI model is usually channel characteristics, and the AI model can captures dynamic changes of input so as to achieve channel inference at output.

The AI channel model can be generally divided into generative and discriminative models. Generative models aim to model the joint probability distribution of input features and output labels. They can be used to model the entire distribution of input (transmitted signal) and output (received signal) variables. Generative models are particularly useful if the goal of channel modeling task is to generate a large amount of synthetic/simulated channel data, when a sufficiently large and heterogeneous input dataset is available. Discriminative models, on the other hand, model the conditional probability of output labels given the input features. They can be used directly model the relationship between input (training) and output channels, without explicitly modeling the entire distribution. Logistic regression and neural networks are typical discriminative models that can be applied to predict channel state or characteristics. They can integrate system parameters with multi-source data and process different data through distinct neural networks. Loss function should be constructed, which can be designed based on feature enhancement according to AI model interpretation, highlighting the impact of particular features. Alternatively, loss function can be driven by prior channel models to construct propagation constraints. Moreover, priori knowledge such as propagation mechanism can be used for model tuning to enhance prediction accuracy and generalization ability. Specifically, the AI model can either use prior knowledge from existing channel models as references, or integrate physical propagation features into the AI network as additional input.

As an illustrative example, we present path loss prediction using both Generative Adversarial Network and Convolutional Neural Network (CNN) with ResNet as backbone, where the information of building map is used to train the network to output path loss. Fig. 1 shows prediction results of AI model, compared with ray tracing (RT) simulations. It can be observed that AI-predicted path loss shows fairly high consistency with simulations, reaching a Root Mean Square Error (RMSE) at 2.6 dB for generative model and 5.4 dB for discriminative model, respectively.

In practice, a combination of generative and discriminative models can be used for channel modeling. For example, a generative model might be used to generate a set of diverse channel conditions, and a discriminative model can be trained on real-world data for specific classification tasks. Ultimately, the choice of which model to use depends on data availability, target use case, and requirements of the channel modeling task, as well as available computational resources. In regards to data requirements, generative models require more data to accurately model the joint distribution, especially if the channel conditions are complex. In terms of the target use case, if the intention is to provide a "black box" for modeling the channel, generative models are a better choice, as they can generate new samples or simulate different (even unseen) scenarios. Finally, in terms of computational resources, discriminative models are computationally more efficient as they focus on the decision boundary without explicitly modeling the entire distribution.

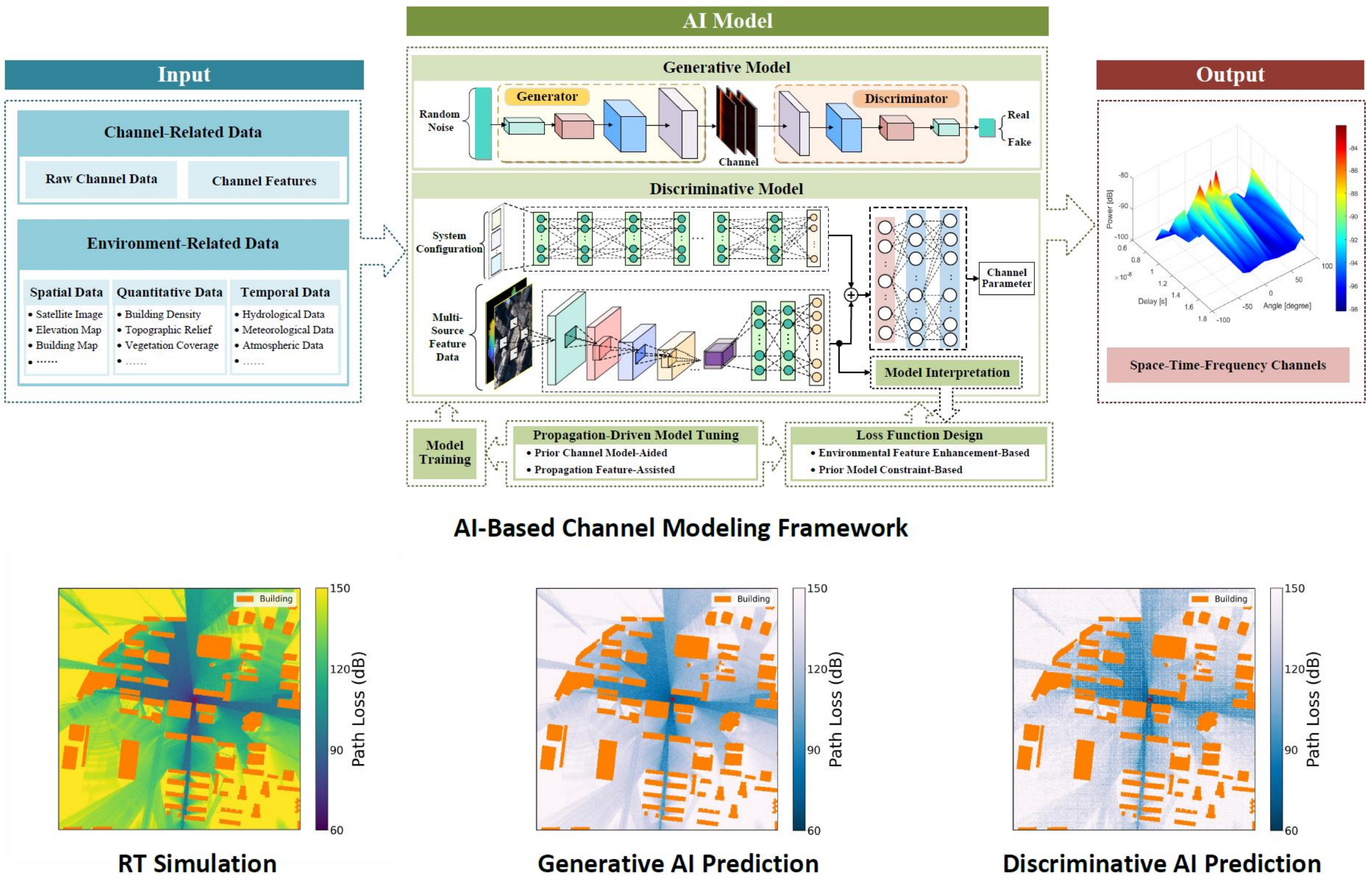

Fig. 1. Illustration of AI-based channel modeling framework (upper), and illustration of AI-based channel modeling performance (bottom). For the illustration of AI-based modeling performance, it is conducted within a 1 km × 1 km area of urban campus scenario (Beijing Jiaotong University, China). A self-designed RT simulation platform is used for comparison, which has been calibrated with measurement data and has a fairly high accuracy (with 2.4 dB RMSE). Here, TX is indicated as the red circle in the middle, carrier frequency is 5.9 GHz, bandwidth is 30 MHz, and TX/RX heights are 15 m and 2 m, respectively.

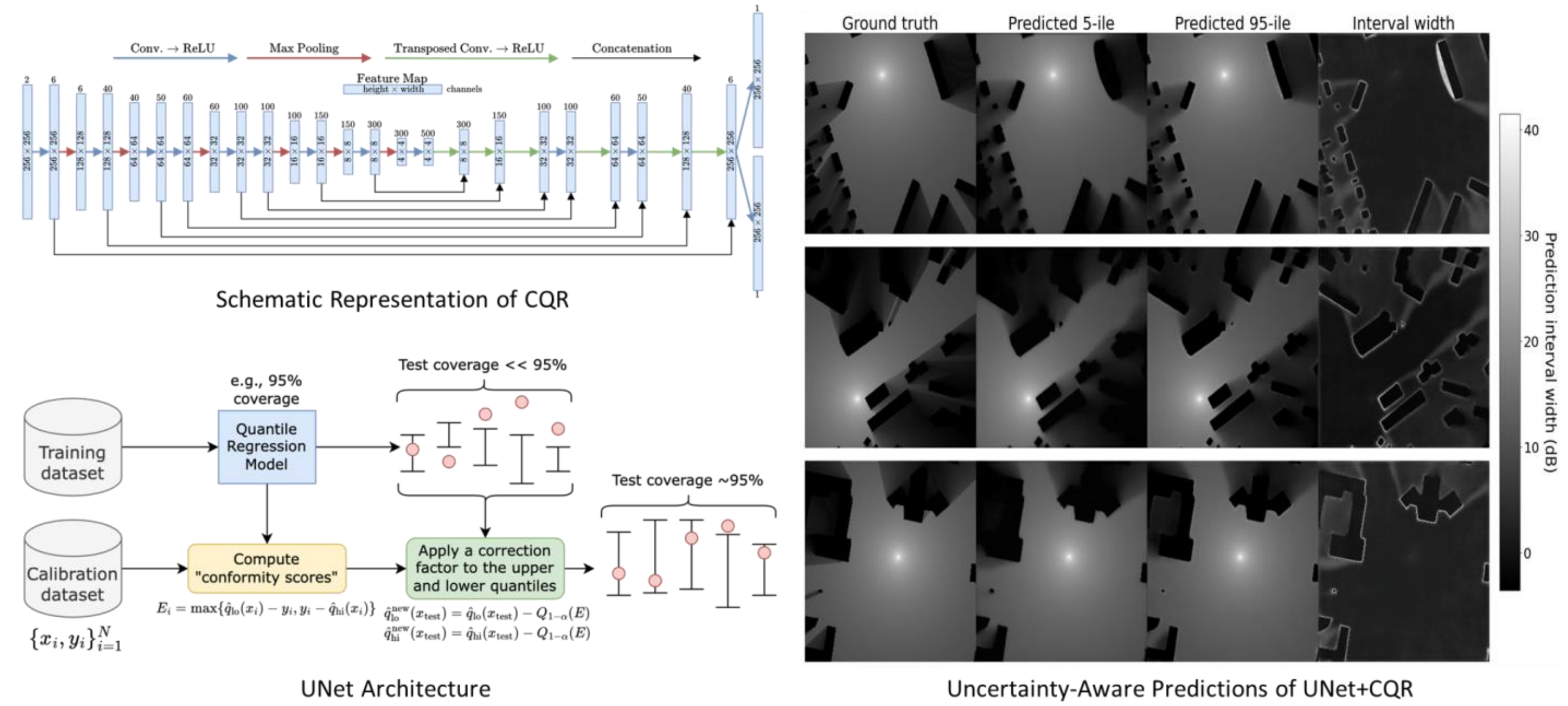

Fig. 2. Illustration of uncertainty-aware AI for channel modeling.

## III. Uncertainty-Aware AI for Channel Modeling

Many of the popular AI models for channel modeling, such as DNNs or decision trees/random forests, output point predictions by default. This is a significant limitation, as it makes the end-users unaware of the potential prediction errors that the model might be making. In other words, although the model achieves a low average error, we generally have no information on the uncertainty associated with individual predictions. For instance, when leveraging AI-based channel models for wireless network planning, informing users that a prediction has high uncertainty allows for more informed and less risky decision-making. Alternatively, the predictive uncertainty can be employed in an Active Learning context, such that the data collection process (e.g., a measurement campaign or RT simulation) is guided towards annotating inputs whose predictions are highly uncertain. It is, therefore, of great importance and practical utility to estimate the predictive uncertainty when leveraging AI for channel modeling.

A natural solution for uncertainty quantification is to provide probabilistic predictions (e.g., in the form of a confidence interval) instead of point predictions. Even though there are many solutions for probabilistic AI, such as Bayesian Neural Networks and Quantile Regression (QR), they do not provide any formal guarantee that their probabilistic predictions will be truthful to the actual data distribution. Recent approaches based on Conformal Prediction (CP) solve this fundamental issue. In contrast to prior methods, CP makes minimal assumptions about the data distribution (i.e., data exchangeability, which is a weaker assumption than independent and identically distributed data) while providing explicit, finite-sample probabilistic guarantees. An attractive and easy-to-apply CP algorithm is Conformal Quantile Regression (CQR) [10]. CQR extends classical QR algorithms by outputting prediction intervals guaranteed to contain the ground-truth with a user-specified probability. The CQR workflow requires a "proper" training set to fit a pair of QR models and a calibration set to perform a post-hoc correction of the predicted quantiles. More specifically, the model predictions on the calibration set are used to compute a "correction factor" to the predicted quantiles, such that the resulting prediction intervals are theoretically guaranteed to contain the ground truth with a user-specified probability. The idea of CQR is to take conventional QR models, such as deep quantile neural networks, and correct their predictions a posteriori. This is because conventional QR models (e.g., deep quantile neural networks) do not guarantee the user-specified confidence interval coverage levels (e.g., 95%). CQR computes "conformity scores" from a held-out calibration set, and then applies a corrective factor to the predictive confidence intervals equal to the 1-α empirical quantile of the calibration set conformity scores. Specifically, for CQR, the conformity score is shaped to be positive if the ground-truth falls outside the predicted interval, and negative otherwise. We then compute a "correction factor" equal to the 1-α empirical quantile of the conformity scores in the calibration set. Finally, to perform a prediction on a new test point, we first obtain a confidence interval from our QR model, and then adjust it by subtracting and adding the correction factor to the lower and upper predicted quantiles, respectively. Remarkably, such intervals are theoretically guaranteed to contain the ground truth with 1-α probability. We illustrate in Fig. 2 a schematic representation of the CQR algorithm, including the computation of the conformity scores and the correction factors.

As an illustrative example, we consider a radio map estimation task based on the RadioMapSeer dataset [11]. This dataset comprises a collection of 2D urban maps alongside path loss estimations derived via RT simulations. Specifically, coverage maps were computed for 800 urban maps with 80 transmitters (TXs) per urban map, using the WinProp simulation software. The objective is to leverage AI to provide path gain estimations that approximate the quality of RT simulations while being much faster to compute. Instead of point predictions, we aim to deliver prediction intervals containing the ground truth path gain values with 90% probability. We consider a UNet model that provides pixel-level predictions given an input 2D map and the TX location [11]. In particular, we train a two-output UNet on the quantile loss to predict the 5-ile and the 95-ile of the path gain. In Fig. 2, we illustrate in detail the architecture of the UNet, annotating for each layer the dimensionalities of the computed feature maps. The model interleaves convolution and downsampling in the encoding part, followed by transposed convolutions in the decoding part. The top and bottom output represent per-pixel predictions of the bottom and upper quantiles, respectively, of the path gain. The model takes a 2D map of size 256 × 256 as an input, and outputs lower and upper quantile coverage map predictions, both of dimension 256 × 256. The upper and lower quantile predictors share all layers but the last. We then apply CQR to "conformalize" the predicted quantiles to achieve the desired coverage guarantee. In Fig. 2, we display three illustrative uncertainty-aware predictions of UNet+CQR. We show separately the lower and upper bounds of the predictive intervals and their width. The path loss values are normalized in grayscale such that 0 and 1 correspond to the maximum and minimum path losses, respectively. Interestingly, we observe that the regions with the highest predictive uncertainty are primarily on the Line-of-Sight (LoS)/Non-Line-of-Sight (NLoS) boundary and secondarily in the NLoS region. This provides a clear, practical indication of where the model predictions become more unreliable. On average, in our test set, we observe that the vanilla UNet provides intervals with 93.0% coverage and 5.19 dB width, displaying a systematic overestimation of the predictive uncertainty. In contrast, the intervals from UNet+CQR provide, on average, 90.7% coverage (almost equal to the theoretically expected value) with a 4.53 dB width, outperforming the vanilla UNet in both calibration and sharpness. Overall, we conclude that CP, particularly CQR, stands as a reliable and theoretically backed solution for uncertainty-aware AI-based channel modeling.

## IV. Physics-Informed AI for Channel Modeling

A fundamental issue of AI-based channel modeling is the generalization to scenarios unseen at training time. Though channel models are generally site-specific, it is expected that an AI-based channel model will be able to generalize to propagation conditions like the ones seen during training. Generalization can be particularly challenging if the training data, e.g., samples from a measurement campaign, is scarce. AI models are generally unaware that the training data stems from physical phenomena, which makes them prone to overfitting in data-scarce scenarios. A solution is incorporating domain knowledge from the physics domain into the AI pipeline. Here, following prior literature, we taxonomize physics-informed AI for channel modeling as follows: a) augmenting input data to AI model with physics-based features, b) leveraging AI model architectures with "inductive biases" suitable for the considered channel modeling problem, and c) embedding a priori known physical laws in the training loss of AI model.

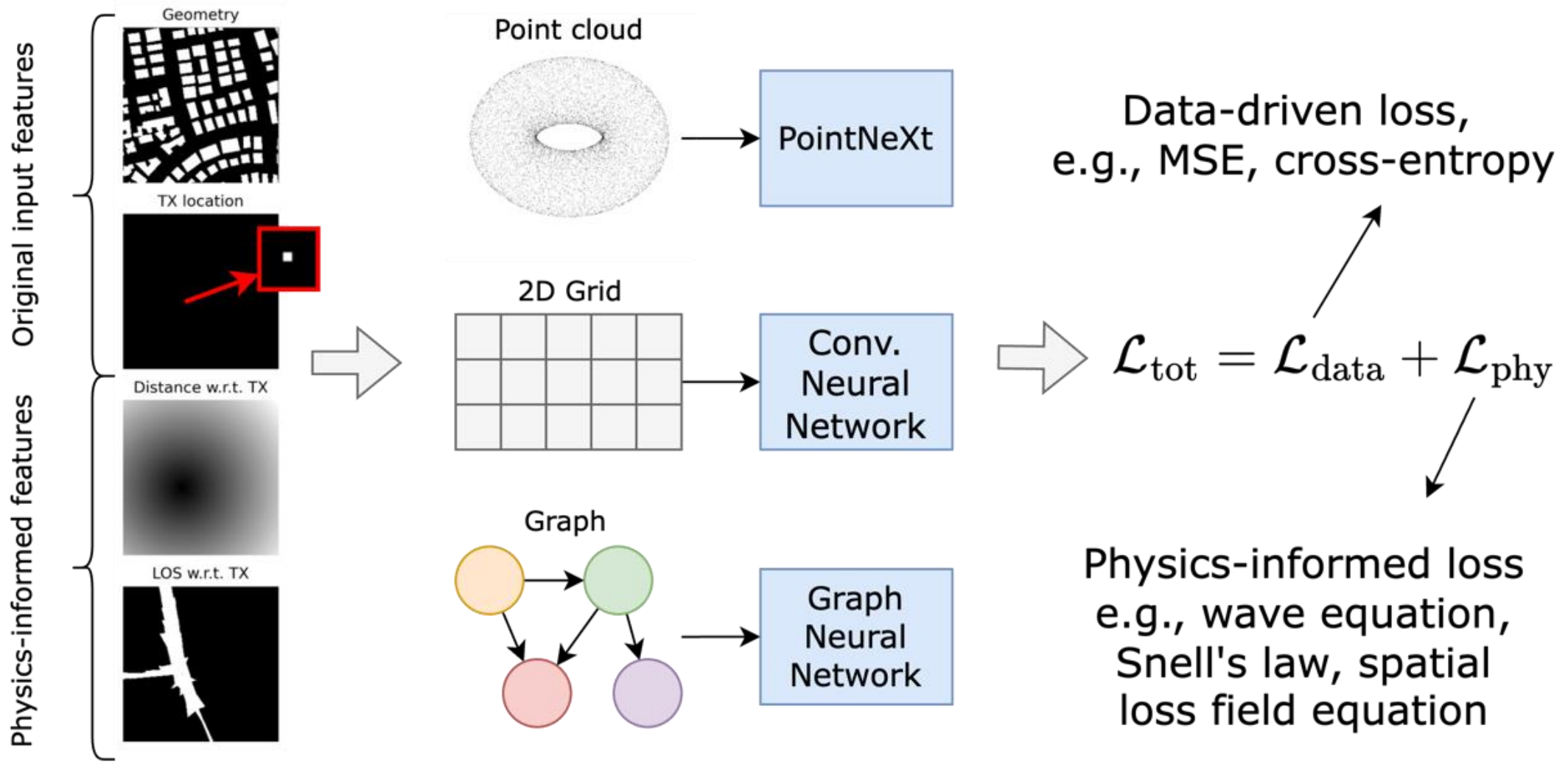

Fig. 3. Summary of physics-informed AI techniques for channel modeling: Physics-informed data augmentation (left). Physics-informed model architectures (middle). Physics-informed loss functions (right).

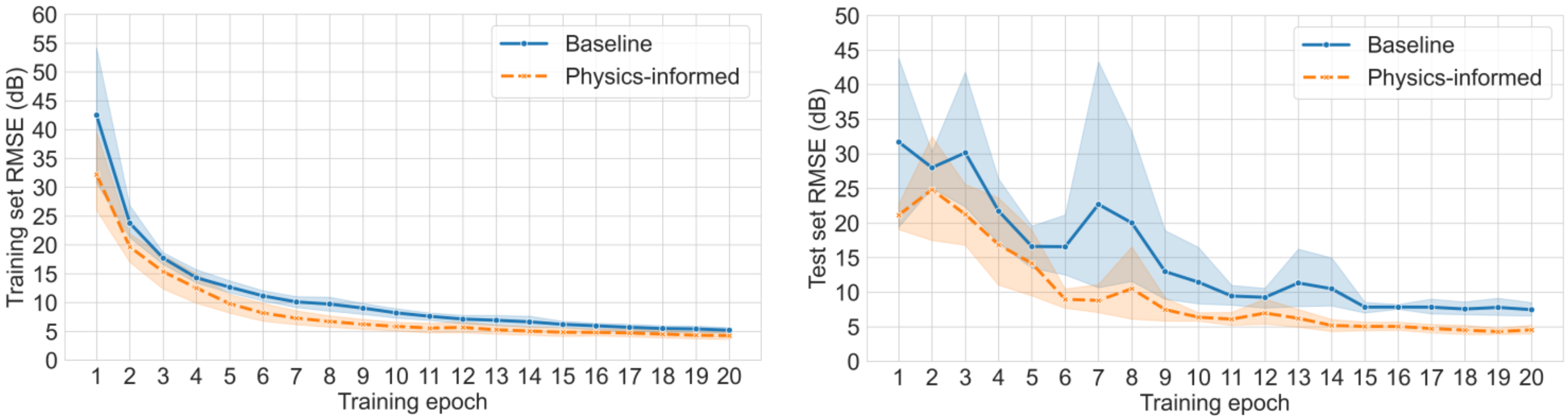

Fig. 4. Training set (left) and test set (right) RMSEs of the baseline vs. the physics-informed UNets for path loss regression.

**Physics-Informed Data Augmentation.** This consists of engineering physics-informed features that can boost the model's predictive power. The idea is to represent known, possibly non-linear, interactions between input and output that would have otherwise needed to be learned from data. As an illustrative example, consider a path loss regression problem given environmental features, e.g., building density, etc. On top of these features, it can be highly beneficial to specify, for instance, free-space (or a more specialized statistical propagation model) path loss value, LoS condition, or the presence of a diffracting edge near the receiver (RX). As an illustrative example, in Fig. 3 we illustrate the original features from the UNet dataset (i.e., environment geometry and TX position), and two potential physics-informed features, namely, the per-pixel distance between the TX and the RX, and the per-pixel LoS condition.

**Physics-Informed Model Architectures.** While exhibiting exceptional function approximation capabilities, AI models such as multi-layer perceptrons or boosted trees are unaware of the physical structure underlying the input data. Geometric Deep Learning models such as CNNs are particularly promising architectures for incorporating physical knowledge into AI-based channel models. This is because the computational model of Geometric Deep Learning implicitly encodes useful “inductive biases” [12], e.g., the data represents a 2D/3D Euclidean space, a graph structure, etc. In Fig. 3 we illustrate three data modalities relevant to wireless channel modeling, namely, point clouds, grids/meshes and graphs, alongside ML models tailored for processing these data types directly.

**Physics-Informed Loss Functions.** The idea is to shape the loss function so that physically inconsistent predictions are penalized. In practical terms, training is performed by minimizing a standard error function plus a “physics-informed loss” that incorporates some a priori known physical laws. The only constraint is that the physics-informed loss must be compatible with the training algorithm of the considered AI model, e.g., stochastic gradient descent for neural networks. For instance, one could incorporate differential equations from electromagnetic theory or laws from geometrical optics. In Fig. 3 we illustrate the decomposition of the total loss into a data-driven loss and a physics-informed loss, and we report some illustrative examples.

As an illustrative example of physics-based AI for channel modeling, we consider again a radio map estimation task on the RadioMapSeer dataset [11]. As before, we train a UNet to predict the path loss given an input map and TX location. UNets, being CNNs, incorporate physics-based inductive biases since they assume that the input lies in a 2D grid sampled from an Euclidean space. To perform physics-informed data augmentation, we add the following inputs in the form of additional channels: a) 3D distance from TX, and b) LoS condition with respect to TX. Note that, the 3D distance feature map subsumes TX position while providing richer spatial information. We consider less than 1000 radio coverage maps among the first 700 urban maps in the RadioMapSeer dataset for training to simulate a data-scarce scenario, and we keep the last 100 urban maps for testing. In Fig. 4, we plot the training and test set RMSE of the baseline and of our physics-informed model. We observe that, even though the baseline and physics-informed models achieve similar training errors, the physics-informed model achieves a lower test error, indicating that it generalizes better than the baseline given the same training budget. Specifically, by including these simple physics-based features, we improve the test set RMSE, on average, from 7.47 dB to 4.54 dB, i.e., a 39% improvement.

## V. Interpretable AI for Channel Modeling

Even though AI models such as DNNs can learn exceptionally complex functions from data, they, unfortunately, provide limited insight into the actual reasoning behind the model predictions. This lack of interpretability becomes a severe drawback in scenarios where understanding the physical layer characteristics of wireless channels is crucial. Because of this, a network operator might prefer relying on more straightforward, interpretable, and battle-tested models to black-box AI-based solutions. To solve this fundamental issue, eXplainable AI (XAI) methodologies are proposed to enhance the transparency and trustworthiness of black-box AI models [13]. XAI methods allow for deconstructing and analyzing complex AI models, making their internal mechanics and decision-making processes understandable. Though there are many different angles for XAI, we here focus on what we believe is one of the most promising methodologies for interpretable channel modeling, that is, high-performance symbolic regression.

Symbolic regression is fundamentally a supervised learning task. The goal of symbolic regression is to learn compact analytical expressions from a dataset of input-output exemplars. Mathematical expressions can be fit either to raw measurements data, or to predictions from a black-box ML model. In the former case, our objective is to discover via symbolic expressions the physical laws hidden in the data. In the latter case, known in the literature as symbolic distillation, our objective is to approximate the function learned by an ML model with a simple formula. A state-of-the-art implementation of modern symbolic regression is PySR [14]. In PySR, symbolic expressions are constructed as combinations of a) atomic operators applied to the input features, such as summations, multiplications, exponentiations, differential/integral operations, etc., and b) arbitrarily valued real constants. Fig. 5 illustrates a simple expression tree fit to the Friis' formula, comprising unary and binary operators. This illustrative expression tree presents a clearly spurious term in the form of a cosine applied to the signal frequency. Symbolic regression makes it easier to detect implausible input-output relations, as opposed to conventional black-box Machine Learning models such as DNNs. Nevertheless, it is immediately apparent that, even for a few candidate inputs and operators, the combinatorial explosion of the search space makes brute-force search unfeasible. To make the problem computationally tractable, PySR implements massively parallelized evolutionary algorithms that display excellent scalability with respect to the amount of available computing. Moreover, since simpler analytics expressions are generally preferable, the exploration process can be "guided" by constraining the complexity of the generated expressions, for instance, by forbidding nesting of sines and cosines or by limiting the number of symbols that can appear as an argument to an operator, or by excluding functional forms deemed to be implausible (e.g., based on prior knowledge of the considered physical domain). Such constraints are purely knowledge-driven and translate to injecting useful inductive biases in the learning process of the model, in a similar fashion to Physics-Informed AI. However, symbolic regression differs from Physics-Informed AI since its primary goal is not to improve a ML model's performance in a task leveraging domain knowledge of the physical domain, but to provide a set of interpretable mathematical formulas to the user. In the end, the algorithm will return a Pareto front of candidate symbolic expressions, each achieving a different trade-off regarding the loss function and complexity (by default defined as the total number of variables, operators, and constants).

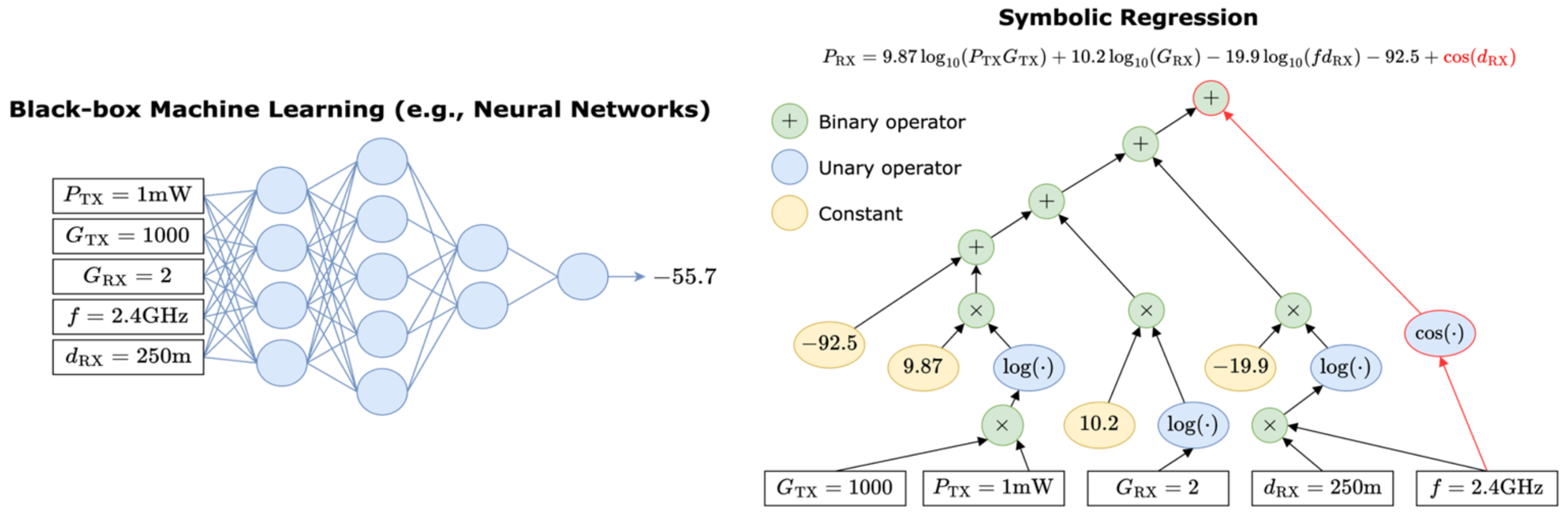

Fig. 5. Black-box AI (left) vs. Symbolic regression (right) for a simple free-space path loss regression task. Conventional black-box approaches provide no insight into the relationship between the predicted output and the input features. In contrast, symbolic regression algorithms produce a compact analytical expression through a symbolic tree, which allows domain-expert inspection and identification of biases (highlighted in red).

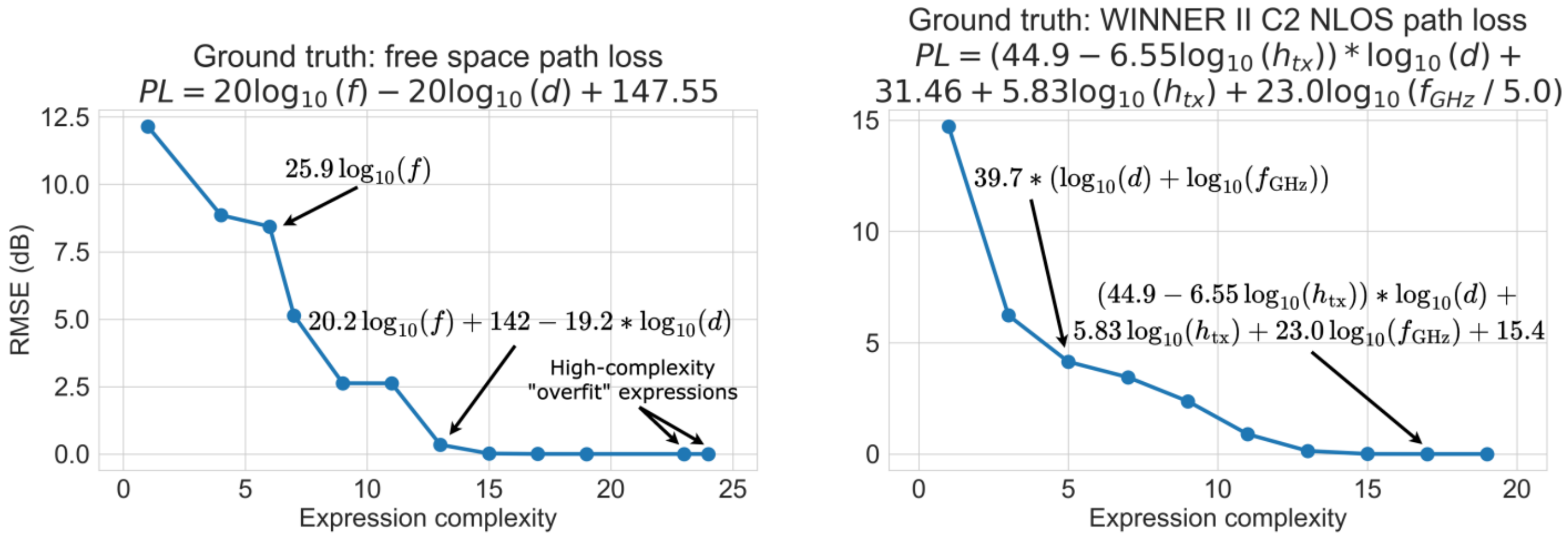

Fig. 6. RMSE vs. complexity (i.e., total number of operators, variables, and constants) of the best-found symbolic expressions by PySR for free-space path loss and the WINNER II C2 NLoS channel model.

In the context of wireless channel modeling, symbolic regression can be leveraged to learn, e.g., path loss models from a massive measurement campaign, in a similar fashion in which empirical channel models are developed. Instead of fitting the parameters of standard "blueprint" formulas, e.g., Hata-like models, symbolic regression can provide simple closed-form expressions potentially capturing complex propagation phenomena. The generated symbolic expressions can then be transparently analyzed, evaluating different trade-offs (e.g., asymptotic behavior, generalizability, and goodness-of-fit versus expression complexity) and possibly corrected post-hoc by leveraging domain knowledge.

As a motivating example, we apply symbolic regression to recover well-known analytic channel models from data. In particular, we consider "discovering" the free space path loss formula as a simple example, and the WINNER II path loss model for urban macrocells (scenario C2 in [15]) as a more involved example. For both tasks, we generate noiseless synthetic datasets consisting of 1000 exemplars. We consider addition, multiplication, division, and power elevation as binary operators, and base-10 logarithm, sine, cosine, and squaring as unary operators. Though not every operator necessarily appears in the ground-truth expression, we include on-purpose "confounding" operators to increase the difficulty of discovering the law hidden in the data. We nevertheless applied some commonsense constraints, such as excluding redundant combinations of operators (e.g., products inside a logarithm, since a sum of logarithms is equivalent) and deep nesting of unary operators. In practice, as mentioned before, the end-user will apply further domain knowledge to filter out implausible combinations of operators. In both scenarios, the algorithm discovers the ground-truth functional form and the hidden physical constants. For the free space path loss formula, we consider frequency and distance from TX to RX as inputs. For the WINNER II channel model, we additionally consider the TX antenna height. In Fig. 6, we plot the RMSE in dB as a function of the total number of symbols in the discovered symbolic expressions. In both cases, we observe that the error vanishes to zero, i.e., the algorithm recovers the correct functional form and correctly approximates the hidden numerical constants. Upon visual inspection of the produced expressions, we observe that the algorithm proposed, alongside the correct functional forms, more complex expressions achieving similar RMSEs, but comprising spurious sine and cosine terms, similar to what is illustrated in Fig. 5. This is a case of "overfitting," which symbolic regression makes it easy to detect, as opposed to conventional, inscrutable AI black boxes. On top of post-hoc inspection of the discovered expressions, a domain-expert user can prevent the occurrence of spurious terms by forbidding or imposing high costs to implausible functional forms. For the WINNER formula, note that the algorithm discovers a different additive numerical constant of 15.4 as opposed to the ground truth of 31.46. This is, actually, a correct result: the symbolic regression algorithm discovers the difference between 31.46 and -10log105 (i.e., the constant term that divides the frequency). This is because a single constant term has lower complexity with respect to a constant minus a logarithm of another constant. This simple result highlights the capability of modern symbolic regression to find compact and accurate analytical expressions for channel modeling.

To conclude, we underline that, in general, there is a trade-off between an ML model's performance and its interpretability. In our case, symbolic regression models are significantly more complex to train than conventional ML models and might not scale well if the number of input features grows very large (e.g., more than a hundred). XAI algorithms for conventional ML models do exist, e.g., [13], but they provide a limited amount of information on the learned input-output mapping. How to balance this trade-off will depend on the operational requirements of specific use cases, i.e., whether black-box channel models are acceptable, or some degree of interpretability is mandatory.

## VI. Conclusion

This paper presents a vision of AI-based channel modeling, which will be a new paradigm for future channel modeling and prediction. The deep integration of AI and radio propagation theories empowers a high-precise and generalized channel modeling framework. Based on the proposed framework, we further explore and highlight several specific technical topics: how to characterize and optimize the uncertainty of model outputs, how to strengthen the impacts of physical propagation characteristics, and how to enhance the interpretability of models. We validate the potential of AI-based channel modeling through a series of illustrative numerical results.